# Adding Gauge Fields to Kaplan's Fermions

T. Blum and Leo Kärkkäinen

Department of Physics, University of Arizona
Building 81, Tucson, AZ 85721, USA

We experiment with adding dynamical gauge field to Kaplan (defect) fermions. In the case of U(1) gauge theory we use an inhomogenous Higgs mechanism to restrict the 3d gauge dynamics to a planar 2d defect. In our simulations the 3d theory produce the correct 2d gauge dynamics. We measure fermion propagators with dynamical gauge fields. They posses the correct chiral structure. The fermions at the boundary of the support of the gauge field (waveguide) are non-chiral, and have a mass two times heavier than the chiral modes. Moreover, these modes cannot be excited by a source at the defect; implying that they are dynamically decoupled. We have also checked that the anomaly relation is fullfilled for the case of a smooth external gauge field.

## 1. INTRODUCTION

It is difficult to implement chiral fermions on the lattice. The Nielsen-Ninomiya theorem forbids a local, hermitian and translation-invariant action for chiral fermions [1]. A possible way to circumvent this was proposed by Kaplan [2]. By considering the even d-dimensional theory as a low energy limit of an odd d+1-dimensional vector like theory at a given mass defect, he showed that the free theory could be rendered d-dimensional and chiral.

We propose to restrict the gauge fields with a similar mechanism: by adding a Higgs doublet to the theory with an inhomogenous mass term, the gauge symmetry is spontaneously broken outside the defect, which confines the lower dimensional gauge fields to the defect. The fermions were already confined by their own inhomogenous mass term.

As a proof of concept we perform simulations of pure U(1) gauge theory with a Higgs doublet in 3 dimensions and show the U(1) fields are confined to a 2d plane. By carefully choosing the type of the defect and the Higgs field we show by means of MC simulation that the coupled Higgs-gauge system produces the correct 2d (analytically known [3]) U(1)-gauge model at the defect [4]. Furthermore, we show that in the presence of these gauge fields the fermions retain their chiral nature.

## 2. PROPOSAL

We begin with the continuum Lagrangian density for electrodynamics with scalar and fermion fields:

$$\mathcal{L} = \partial_\mu \sigma \partial^\mu \sigma \ + \ (m_H{}^2 - g_\sigma \sigma^2)\phi^* \phi + g_0(\phi^* \phi)^2$$
$$+ D_\mu \phi^* D^\mu \phi \ + \ F_{\mu\nu} F^{\mu\nu} + \bar{\psi}(\gamma^\mu D_\mu + \sigma)\psi, \quad (1)$$

with $\phi$ a charged scalar (Higgs field), $\psi$ a Dirac fermion and $\sigma$ a neutral scalar with coupling $g_\sigma$. $m_H$ is the Higgs mass and $g_0$ its self-coupling. Here, $-\sigma^2$ is essentially a negative mass term for the Higgs field. The $\sigma$ field couples to the Fermions and to the Higgs field. It can be spontaneously broken to positive or negative values. In particular, it is possible to create a $d$-dimensional domain wall between differently ordered volumes of the d+1-dimensional space. At a place where $< \sigma >$ changes sign, Kaplan's inhomogeneous mass term is formed dynamically. If we let $\sigma \to \pm\infty$ outside the defect ($\sigma = 0$ defines the defect) then $\phi$ acquires a vacuum expectation value and the photon becomes massive. If $m_H$ is also very large (but $m_H{}^2 < g_\sigma \sigma^2$), then on the defect the Higgs field decouples and we are left with an interacting gauge theory of chiral fermions. The fermionic part of the model is due to Kaplan [2].

Following Kaplan, we name the coordinates $(s, x, t)$, where $s$ denotes the extra dimension. For simplicity, the $\sigma$ field is fixed in the form of a planar order-order interface from positive to negative values perpendicular to the $s$ direction. Then, the coupling $g_\sigma$ can be thought as being rescaled to the $\sigma$ field itself and can be omitted in what follows.

For the fermions this produces Kaplan's chiral fermions at the interface with the $\sigma$ field as the mass defect. For the gauge fields the $\sigma$ field generates the spontaneous symmetry breaking potential for the Higgs fields that render the photons massive outside the defect.

## 3. GAUGE DYNAMICS

The Higgs field can be rotated to the real axis by a gauge transformation which leaves the discretized action $S$ corresponding to the Higgs gauge sector of Lagrangian Eq. (1) in the simple form

$$S = \sum_{(ij)} \kappa(s)\cos(Q\alpha_{ij}) + \beta\sum_P \cos(\alpha_P). \quad (2)$$

It describes a dynamical gauge field with explicit gauge symmetry breaking field. Here $(ij)$ denotes links connecting site $i$ to nearest neighbors $j$ and $Q$ is the charge of the Higgs field. With our lattice size, $16 \times 8^2$, the defect was set at $s = 8$. At the defect the hopping parameter of the Higgs field $\kappa(s=8) = 0$.

In the broken phase, with large $\kappa$ values, a Higgs field with charge $Q = 2$ restricts the gauge fields to a $Z_2$ gauge symmetry, where only $\alpha_i(n) = 0$ or $\pi$ are allowed[6–8].

Consider a system that has a domain wall between the $\alpha = 0$ and $\alpha = \pi$ vacua. This domain wall can be forced onto the system by using twisted boundary conditions for the gauge fields. The domain wall is dynamically pinned to the mass defect.

The simulations are further simplified by not treating the $s$-direction gauge fields dynamically, which are fixed to zero [5]. In the limit $\kappa \to \infty$, this amounts to fixing the remaining $Z_2$ gauge symmetry. This also stabilizes the domain wall to the defect.

At the defect, we can split the action $S$ into two terms, the plaquettes lying in the plane of the defect (the 2d gauge action) and the plaquettes lying out of the plane:

$$S_{s=8} = \beta\sum_P^{2d}\cos(\alpha_P) + \beta(\sum_{ij}\cos(\alpha_{ij}+0) + \\ \sum_{ij}\cos(\alpha_{ij}+\pi)) = \beta\sum_P^{2d}\cos(\alpha_P), \quad (3)$$

where it is assumed the links are frozen to $\alpha = 0$ on the left of the defect and to $\alpha = \pi$ on the right. The unwanted plaquettes in the $s$ direction cancel! The effect of the boundary condition is to decouple the 2d gauge invariant system from the higher dimensional space in which it is embedded.

What about gauge invariance? The 3d gauge symmetry is broken. Consider a transformation, in which the links $\alpha_\mu(n)$ on the defect transform like

$$\begin{cases} \alpha_x(n) \to \alpha_x(n) + \theta(n) - \theta(n+\hat{x}) \\ \alpha_t(n) \to \alpha_t(n) + \theta(n) - \theta(n+\hat{t}) \\ \alpha_s(n) = 0 \to \alpha_s(n) = 0 \end{cases}, \quad (4)$$

where the $\theta(n)$ live at the sites $n$ of the defect. This is not a symmetry of the action, since it changes the values of plaquettes lying out of the defect. But, since these plaquettes cancel eachother, it turns into a 2d gauge symmetry at the defect. This dynamically generated 2d gauge symmetry is the crucial point that allows us to circumvent the no-go theorem. Our simulation[4] of the 3d system gives the exact 2d U(1) gauge theory result[3] for the plaquette on the defect. The result is shown in Fig. 1.

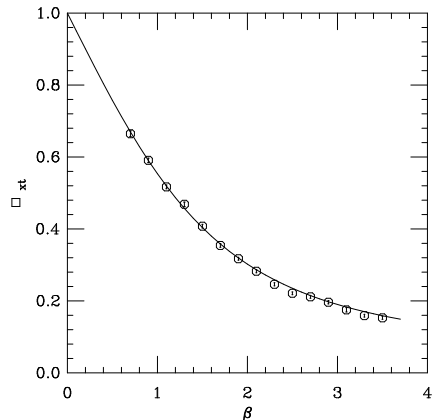

Figure 1. The plaquette on the $4^2 \times 16$ lattice as a function of $\beta$. The solid curve is the exact 2d result.

## 4. FERMION PROPAGATORS

Jansen showed that the correct 2d anomaly is produced if the gauge field is made constant in the $s$-direction [9]. The field was a plane wave in the "physical" $x$ and $t$ directions. However, this way the gauge field is not confined to the defect.

In practice the support(extent) of the chiral mode can be as wide as 6 lattice spacings in the $s$-direction. In order to produce the correct anomaly the gauge field cannot change in the $s$-direction across the support of the fermion. This would mean that the different slices of the same fermion are affected by different gauge fields, which is not desirable.

Because our method produces the gauge field exactly confined at the defect it must be widened to cover the whole support of the fermion. This can be achieved by using a coarser lattice for the gauge fields in the $s$-direction. We only need a finer grid at the defect. The $x$ and $t$ direction dynamical gauge fields at the defect plane are copied to the right and left finer grid of fermions. The lattice structure is shown in Fig. 2. The alternative – to reduce the support of the fermions – does not work, since there is an upper limit for the off-defect fermion mass (in lattice units) for chiral modes. The currents using a coarse lattice for the gauge fields produce the correct anomaly with smooth external gauge fields. The Goldstone-Wilczeck fermion current in the $s-t$ plane with a smooth external gauge field flows out of the defect $s = 8$ to $s = 12$.(Fig. 2). No current flows to smaller $s$ values. The correct anomaly is produced when summed over the support of the chiral mode.

The fermion propagator $G$ is a 2 by 2-matrix

$$G = \begin{bmatrix} -i\psi_R^*\psi_L, -i\psi_R^*\psi_R \\ i\psi_L^*\psi_L, i\psi_L^*\psi_R \end{bmatrix} \qquad (5)$$

For right handed chiral modes only the upper right corner is non-zero, for left handed only the lower left. We measured $G$ by inverting the fermion matrix on 100 dynamical gauge configu-rations with $\beta = 2.5$, $\kappa = 0$ for $s = 8$, and $\kappa = 80$ elsewhere. The picture is qualitatively the same as with free Kaplan fermions. The chirality is preserved even in the presence of dynamical gauge fields.

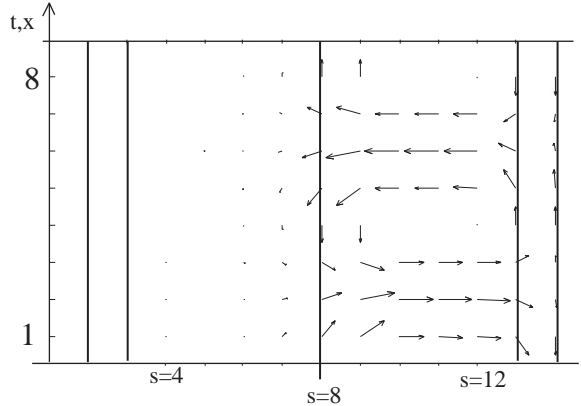

Figure 2. The lattice structure and chiral current. The links between $s = 4$ and $s = 12$ inherit the gauge fields from $s = 8$. The arrows show the current with an external gauge field.

Recently a similar attempt was made by [10]. They restricted the gauge fields to a waveguide by hand and coupled them to Higgs fields at the boundary to preserve 3d gauge invariance. They observed mirrorfermions forming at the boundary of the waveguide. These destroyed the chiral structure of the propagator. Their calculations used only external gauge fields.

We measured the propagator off the defect. Whether the chiral defect fermion is coupled to fermions of the opposite chirality can be checked by measuring the propagator off the defect, when the source is set at the defect. If the formation of mirror fermions were a problem with dynamical gauge fields we would see the component of the opposite chirality soar at the waveguide boundary. In our simulations $G$ retains its chiral character throughout the waveguide. It is displayed in Fig. 3. The dynamical gauge fields do not couple the opposite chirality fermion fields. By setting the source at the waveguide, we excite both chiralities, but with a mass two times the purely

4

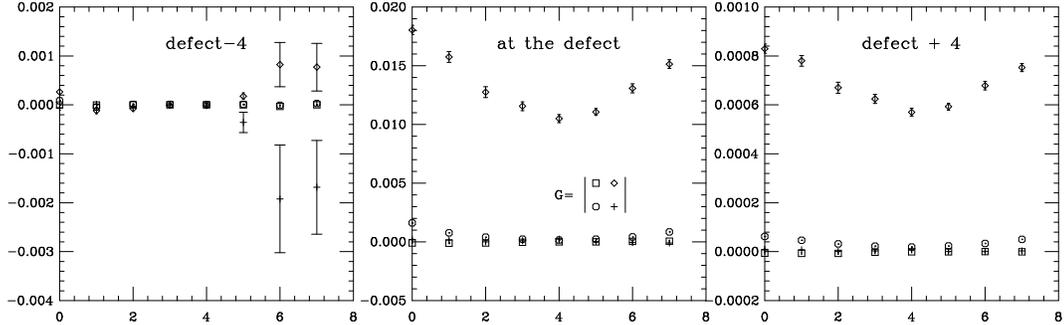

Figure 3. The propagator at the defect $s = 8$, at both boundaries of the waveguide, $s = 4$ and $s = 12$, with source at the defect as a function of $t$ with $x = 4$. The fermion retains its mass and chirality across the whole support. Notice, the propagator at the waveguide boundaries is also very small.

chiral mode, just as one expects with Kaplan's fermions (Fig 4).

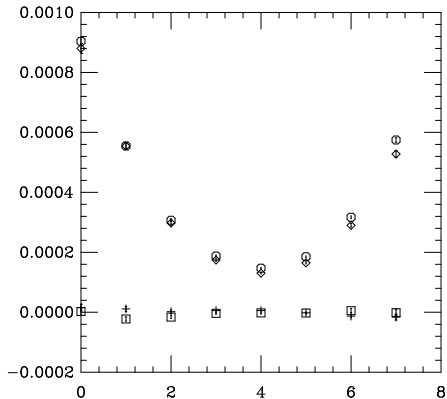

Figure 4. The propagator at the waveguide boundary $s = 12$ with source at the boundary as a function of $t$ with $x = 4$. Both chiralities are excited, but the effective mass is heavy.

## 5. CONCLUSION

We proposed a way to add dynamical gauge fields to Kaplan's fermions. We were able to produce the correct 2d pure gauge dynamics with our system, and the propagators measured with the dynamical gauge fields retained their chiral nature. We found no evidence for mirror fermion formation. The off-defect fermions had a mass twice that of the chiral mode. Also we were able to produce the correct anomaly with an external field using Jansen's method. This work was supported by the U.S. Department of Energy grant number DE-FG02-8SER40213.

## REFERENCES

1. Nielsen and Ninomiya, Nucl. Phys. B185 (1981) 20.
2. D. Kaplan, Phys. Lett. B 288 (1992) 342.
3. J.B. Kogut, Rew. Mod. Phys. 51 (1979) 659.
4. T. Blum and L. Kärkkäinen, *Confining the gauge field to a lower dimensional subspace by an inhomogenous Higgs mechanism*. Arizona University preprint AZ-TH/93-26. Submitted to Phys. Rev. D.
5. J. Distler and Soo-Jong Rey, *3 Into 2 Doesn't Go: (almost) chiral gauge theory in the lattice* PUPT-1386, NSF-93-66, SNUTP 93-27.
6. E. Fradkin and S. Shenker, Phys. Rev. D19 (1979) 3683.
7. M. Creutz, Phys. Rev. D21 (1980) 1006.
8. K. C. Bowler et. al., Phys. Lett. B104 (1981) 418.
9. K. Jansen, Phys. Lett. B288 (1992) 384.
10. M. Golterman et al., *Investigation of the domain wall fermion approach to chiral gauge theories on the lattice* Washington University preprint, UCSD/PTH 93-28.